# Datacom-Agnostic Shortwave QKD for Short-Reach Links

Mariana Ferreira Ramos, Marie-Christine Slater, Michael Hentschel, Martin Achleitner, Hannes Hübel, and Bernhard Schrenk

AIT Austrian Institute of Technology, Center for Digital Safety&Security / Security & Communication Technologies, 1210 Vienna, Austria.

**Author e-mail address**: mariana.ferreira-ramos@ait.ac.at

We investigate the co-existence of 852-nm and 1550-nm QKD with carrier-grade 4×25-Gb/s/λ LAN-WDM over a short-reach interconnect. Shortwave QKD yields a higher key rate and is insensitive to Raman noise, as opposed to 1550-nm QKD.

**Introduction**

The practical quantum key exchange in telecom networks has been well investigated during recent years, including demonstrations of quantum key distribution (QKD) in various field trials. The need for QKD is now enhanced through the rapid scale-up of datacenters, a domain originally characterised by perimeter security but currently on the migration to a zero-trust model where no resource within its short-reach networks is inherently trusted [1]. These short-reach interconnects typically foresee an optical budget in the range of 6 dB [2]. Together with the limited fiber reach, this calls for shortwave QKD schemes. By placing QKD at the border of the visible-light and near-infrared regions, one can build on highly efficient silicon single-photon avalanche photodetectors (SPAD) to eliminate the dead-time limitation of InGaAs SPADs while sustaining key exchange over a lossy channel due to the lower dark count rate. Shortwave QKD has been demonstrated primarily in free-space scenarios [3, 4].

In this work, we experimentally demonstrate datacom-blind QKD operation at 852 nm over a 1-km short-reach optical interconnect featuring a co-existing 100-Gb/s LAN-WDM pipe over four unattenuated classical O-band channels. We show that the QKD performance is insensitive to Raman noise and that a minimum of spectral filters are required to facilitate co-existence. We further investigate the effect of few-mode transmission for the shortwave quantum channel over standard telecom fiber and compare the performance to a 1550-nm QKD layout.

**Shortwave QKD in Co-Existence with Classical 100 Gb/s LAN-WDM in the O-band**

The proposed QKD layout populates the shortwave band at 850 nm to benefit from the wide spectral detuning to the telecom windows in the O- (1310) and C- (1550 nm) bands. This renders it robust under co-existence conditions: Both, the (out-of-band) spontaneous emission tails of classical signals and their more critical (in-band) Raman noise induced during transmission along filled-core fibers fall below a significant level at a detuning beyond 100 THz. Although Raman noise can be mitigated for hollow-core fiber [5], these are hardly available for multi-band systems.

The apparent advantages of shortwave QKD might be offset by two practical challenges: (i) the increased transmission loss of 1.8 dB/km and (ii) the few-mode transmission when using standard single-mode fiber (SMF) as transmission medium, typically having a cut-off wavelength of 1260 nm.

Although a heterogeneous fiber array with a dedicated single-mode core for 850-nm QKD could be used, the support of WDM-enabled co-existence in fiber-scarce optical interconnects with Tb/s/core capacities would dictate the exclusive use of C-band SMF.

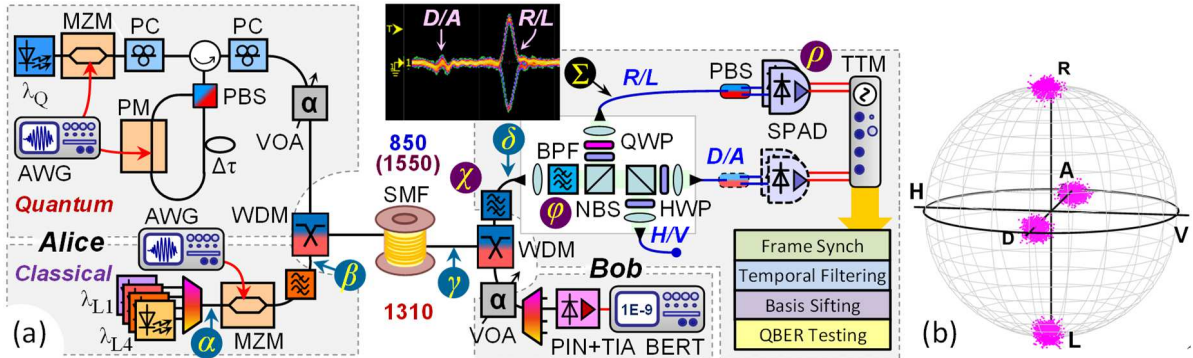

Figure 1: (a) Experimental setup for 852/1550 nm QKD and (b) transmitted states on the Poincare sphere.

Figure 1a presents the experimental setup for investigating shortwave QKD in such a regime. Alice encodes quantum states in the A/D and R/L polarization bases (Fig. 1b) at a wavelength of $\lambda_Q$ = 852 nm using a Sagnac configuration with counter-propagating pulses generated with a Mach-Zehnder modulator (MZM) employed as pulse carver. In this scheme, a polarization state is encoded by phase-modulating a well-defined 45° input state of polarization, which results in a time-multiplexed 800-ps pulse obtained through a short delay $\Delta\tau$ in the asymmetric Sagnac loop. This yields a pair of pulses at a repetition rate of 890 MHz traversing the phase modulator (PM) within the loop in travelling- and anti-travelling wave direction [6]. The resulting symbol rate for two consecutively encoded polarization states is then 445 MHz. A variable optical attenuator (VOA) sets a single-photon launch power with an average photon number of $\mu$ = 0.1 hv/sym. Bob decodes the polarization states and measures them in the A/D and R/L bases, using silicon SPADs with a detection efficiency of 10% and a dark-count rate of 100 cts/s. The counts are registered by a time-tagging module (TTM) and forwarded to a real-time processing stack performing frame synchronization, temporal filtering at 20% of the symbol width and basis sifting before QBER estimation.

| Channel | $\lambda_Q$ [nm] | SMF (MFD) | Data ch. |
|---|---|---|---|
| few-mode shortwave | 852 | SMF-28 (9.2 µm) | LAN-WDM |
| single- mode shortwave | 852 | SM630 (4.3 µm) | no support |
| single-mode C-band | 1550 | SMF-28 (9.2 µm) | LAN-WDM |

Table 1: Investigated QKD layouts.

The robustness of the shortwave QKD system was proven by appending four classical LAN-WDM channels ranging from 1295.56 ($\lambda_{L1}$) to 1309.14 nm ($\lambda_{L4}$), featuring a launch power of 3 and -7 dBm/$\lambda$ after 25 Gb/s/$\lambda$ PAM4 data modulation for the 850- and 1550-nm QKD layout, respectively. Bob tests the BER of these channels using a PIN+TIA receiver. Quantum ($\lambda_Q$) and classical channels ($\lambda_L$) are (de)multiplexed at Alice and Bob using co-existence elements. These filter elements include (i) one cleaning filter for the spontaneous emission tails, which is realized through the interferometric 850/1310 waveband multiplexer at Alice, and (ii) an 850/1310 waveband demultiplexer, and a cascade of filter-WDM 850/1310 multiplexer and a free-space bandpass filter (BPF) with a FWHM bandwidth of 7 nm centered at 850 nm, integrated within Bob's polarimeter, as Raman noise suppression filters.

The short-reach link between Alice and Bob involves a 1-km long ITU-T G.652.B compatible SMF. This makes the shortwave QKD channel at 852 nm subject to few-mode impairments, as will be discussed shortly. Comparison is further made to a C-band QKD system operating at $\lambda_Q$ = 1550 nm, which benefits from single-mode transmission at a performance trade-off with a strongly increased detector dead-time (25 µs), a higher dark count rate (570 cts/s) and its sensitivity to Raman noise arising due to the spectrally closer classical O-band channels.

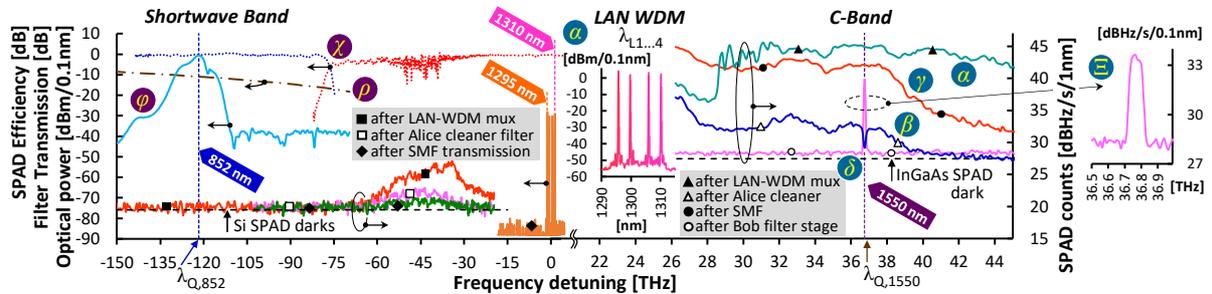

*Figure 2: Spectra in shortwave and C-band after 6-dB of insertion loss for single-photon spectroscopy, transmission of filters and SPAD efficiency.*

**QKD Performance in Shortwave Band and Comparison to C-band QKD Scheme**

Figure 2 reports the optical spectra obtained through a classical and a single-photon spectrometer, relative to the center of the LAN-WDM band (α). At the C-band we see the unfiltered spontaneous emission tails (▲) directly after the LAN-WDM multiplexer (point α in Fig. 1a), highlighting the poor "cleaning" effect of the multiplexer. This function is instead accomplished by the co-existence filter stack at Alice (β), adding a notch at the 1550-nm quantum channel (△). After transmission over 1 km of SMF (γ), Raman noise contaminates the entire C-band (●). It can be widely cleared out at Bob (○), but remains as in-band noise in the 100-GHz quantum channel (Ξ in Fig. 2).

At the shortwave band conditions are by far more favorable. First of all, the detuning of the quantum channel at 852 nm to the LAN-WDM band is much larger. The spontaneous emission tails (■) only show a signature above 1000 nm (-70 THz), even though only one 850/1310 filter-WDM has been used to reject them (□). This is also related to the responsivity roll-off of the SPAD (ρ), which supports their suppression. More importantly, the Raman noise contribution induced by the SMF (⬥) is not stronger than the residual spontaneous emission tails. Consequently, there is no in-band noise contribution at 852 nm, which renders the quantum channel as very robust.

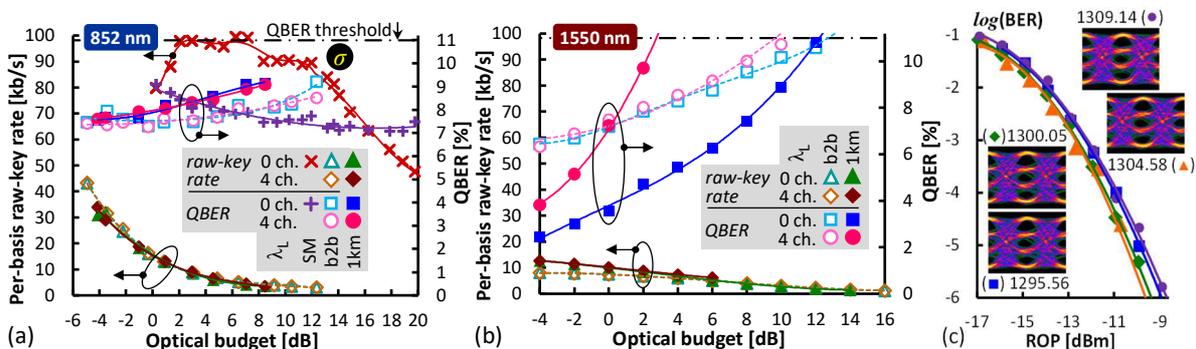

*Figure 3: Raw-key rate and QBER for QKD at (a) 852 nm and (b) 1550 nm. (c) Classical BER performance for 25 Gb/s/λ PAM4 transmission.*

The QKD performance is discussed in Fig. 3 in terms of raw-key rate and QBER, both for the spectral shortwave (Fig. 3a) and C-band layouts (Fig. 3b). Results are presented as function of the optical budget (OB), which is defined as the channel attenuation in excess to a launch level of µ = 0.1 hv/sym. The classical channels ($\lambda_L$) have been kept at a constant power at the receiver, resulting in a worst-case scenario for the quantum channel: It means that besides link attenuation, the OB further represents an increasing crosstalk ratio from the classical to the quantum channel.

For back-to-back shortwave QKD (Fig. 3a) without classical channels, we accomplish a per-basis raw-key rate of 15.6 kb/s (△) at a QBER of 7.3% (□) for an OB of 0 dB. The rather low raw-key rate results from the loss of ~14 dB within the polarimeter due to coupling to single-mode fiber-PBS elements (Σ in Fig. 1a) that additionally act as mode-filtering elements. The raw-key rate is independent of fiber-based transmission (▲) or classical channels co-existing with the QKD signal (◇,◆), whereas the QBER shows a small penalty of 0.73% for few-mode transmission over SMF-28 fiber with a mode-field diameter (MFD) of 9.2 μm (■,●). However, we can sustain QKD operation for an OB of ~10 dB without surpassing the QBER threshold of 11%, despite the presence of four carrier-grade (i.e., unattenuated) classical channels. This proves the excellent robustness of the shortwave QKD scheme to co-existence with classical channels in the O-band telecom waveband, whose typically 50-THz wide Raman tails generated at the transmission fiber does not reach the 850-nm region ($\lambda_Q$) that is 121.7-THz far from the Raman pump ($\lambda_L$).

We further compared the shortwave QKD performance to single-mode transmission over a fiber with a MFD of 4.5 μm while bypassing the free-space polarimeter between the WDM demultiplexer and the fiber-based PBS at Bob. Here, we obtain a much higher raw-key rate of 89.5 kb/s (×) at an OB of 12 dB, which marks the saturation point (σ) of our post-processing stage. This high key rate proves a clear advantage of the shortwave QKD approach by virtue of silicon SPADs operating at a high MHz detection rate without being restricted by a μs-scale dead-time limitation. At the same time, the QBER lowers slightly (+). Nonetheless, this single-mode transmission layout is not compatible with classical-channel co-existence due to the excessively large transmission losses for O-band signals.

In case of the intrinsically single-mode 1550-nm QKD layout, we experience a dead-time limitation for the obtained raw-key rate, which is 7.3 kb/s at an OB of 0 dB for the back-to-back case (△), despite the lower C-band polarimeter loss of 10.2 dB. When the classical channels are present, we see a sharp increase in the QBER, which increases from 3.6% (■) to 7.3% (●) at an OB of 0 dB and transmission over 1 km of SMF. The QBER exceeds the threshold for which a secret key can still be generated at a small OB of 2.8 dB under this co-existence condition. This renders the C-band QKD layout as highly sensitive to the in-band Raman noise of simultaneous data transmission at the 100-GHz quantum channel (Ξ in Fig. 2). We could not clarify the QBER offset between back-to-back (□) and fiber transmission (○), which apparently favors the presence of a SMF.

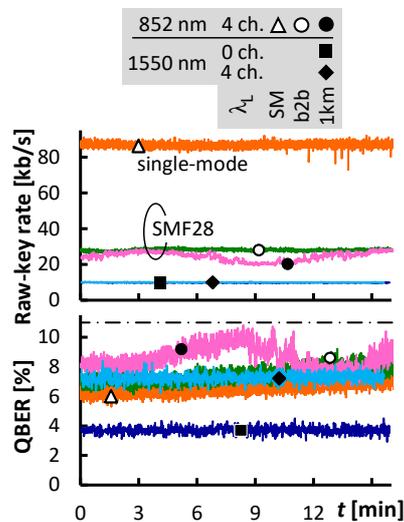

*Figure 4: Longer-term QKD performance.*

Figure 4 reports a longer-term raw-key rate and QBER characterization for all layouts at an OB of 0 dB, except for single-mode shortwave QKD (μ = 0.003 hv/sym, to avoid saturation). A comparison among raw-key rates emphasizes the clear advantage of the shortwave QKD scheme (△,○,●), yielding an up to 86× higher raw-key rate compared to C-band QKD at the same OB. The ringing in the 852-nm rate over SMF channel (●) is attributed to temporal variations for the few-mode coupling. Several layouts perform in a stable manner below the QBER threshold of 11%.

Finally, Fig. 3c reports the BER performance for the classical channels carrying 25 Gb/s/λ PAM4 as a function of the received optical power (ROP). The reception sensitivity is better than -11.1 dBm at the KR4 FEC threshold of $2.2 \times 10^{-4}$. Given the 10-dB attenuated classical signal launch for C-band QKD, a more sensitive APD or SOA+PIN receiver would be required to sustain data transmission at a margin of 3 dB, thus raising the complexity and cost.

**Conclusion**

Data-blind shortwave QKD operation over a 1-km short-reach interconnect was demonstrated under co-existence with 100 Gb/s LAN-WDM over four classical O-band channels. Long-term stability has been accomplished for few-mode shortwave QKD over SMF28 fiber with 9.2 μm MFD, also proving its robustness to Raman noise by achieving a penalty-free raw-key rate of 20.3 kb/s under classical channel co-existence.

**Acknowledgement**

This work has received funding from the EU Horizon Europe Work Programme under project QSNP (no. 101114043) and Qu-Test (no. 101113901).

**References**

[1] S. Sarkar et al., "Security of Zero Trust Networks in Cloud Computing: A Comparative Review," Sustainability 14, 11213 (2022).

[2] X. Zhou et al., "Beyond 1 Tb/s Intra-Data Center Interconnect Technology: IM-DD OR Coherent?," JLT 38, 475 (2019).

[3] G. Vest et al., "Quantum key Distribution with a Hand-Held Sender Unit," Phys. Rev. Applied 18, 024067 (2022).

[4] G. Vallone et al., "Free-space Quantum Key Distribution by Rotation-invariant Twisted Photons," Phys. Rev. Lett. 113, 060503 (2014).

[5] F. Honz et al., "First Demonstration of 25λ × 10 Gb/s C+L Band Classical / DV-QKD Co-Existence Over Single Bidirectional Fiber Link," JLT 41, 3587 (2023).

[6] C. Agnesi et al. "All-fiber Self-compensating Polarization Encoder for Quantum Key Distribution," Optics Letters 44, 2398 (2019).